\documentclass[letterpaper,twocolumn,10pt]{article}
\usepackage{usenix}

\usepackage{tikz}
\usepackage{amsmath}

\usepackage{subfiles}
\usepackage{graphics}
\usepackage{hyphenat}
\usepackage{array}
\usepackage{balance}

\usepackage{caption,tabularx,booktabs}
\usepackage{pifont}
\usepackage[ruled,vlined,linesnumbered]{algorithm2e}
\SetAlFnt{\small}
\usepackage{algpseudocode}
\algnewcommand\Input{\item[\textbf{Input:}]}
   \algnewcommand\Output{\item[\textbf{Output:}]}

\usepackage{xcolor,colortbl}
\usepackage{multirow}
\usepackage{wrapfig}

\newcommand*\halfcirc[1][1ex]{%
  \begin{tikzpicture}
  \draw[fill] (0,0)-- (90:#1) arc (90:270:#1) -- cycle ;
  \draw (0,0) circle (#1);
  \end{tikzpicture}}

\newcommand{\technique}{\textit{Chameleon}}
\newcommand{\identifier}{\textit{\technique{} Identifier}}
\newcommand{\monitor}{\textit{\technique{} Monitor}}
\newcommand{\logger}{\textit{Logger}}

\usepackage{listings}
\usepackage{makecell}
\usepackage{arydshln}
\usepackage{enumerate}
\usepackage[shortlabels]{enumitem}
\usepackage{diagbox}
\usepackage{rotating}
\usepackage{subcaption}
\usepackage{float}
\usepackage{bbding}
\usepackage{caption}
\definecolor{commentsColor}{rgb}{0, 0.5, 0}
\definecolor{keywordsColor}{rgb}{0.000000, 0.000000, 0.635294}
\definecolor{stringColor}{rgb}{0.558215, 0.000000, 0.135316}
\definecolor{applegreen}{rgb}{0.55, 0.71, 0.0}
\lstdefinestyle{cppstyle}{
    language=C++,
    backgroundcolor=\color{gray!10},
    basicstyle=\ttfamily\footnotesize,
    commentstyle=\itshape\color{gray},
    keywordstyle=\color{blue},
    stringstyle=\color{red},
    numbers=left,
    numberstyle=\tiny\ttfamily,
    numbersep=5pt,
    frame=lines,
    framerule=0.4pt,
    breaklines=true,
    tabsize=4,
    showstringspaces=false,
    xleftmargin=10pt,
    xrightmargin=0pt,
    belowskip=0pt,
    aboveskip=0pt,
    lineskip=-0.5pt,
    columns=flexible,
}

\lstdefinelanguage{myLang}
{
	morekeywords={bool, u32_int},
	sensitive=false,
	morecomment=[l]{//},
	morecomment=[s]{/*}{*/},
	morestring=[b]"
}
\lstdefinelanguage{myLangLL}
{
	morekeywords={entry, alloca, \define, i32, i64, call, void},
	sensitive=false,
	morecomment=[l]{//},
	morecomment=[s]{/*}{*/},
	morestring=[b]"
}

\definecolor{greenannoback}{RGB}{230,244,214}

\usepackage{titlesec}
\titleformat{\paragraph}[runin]{\normalfont\normalsize}{}{0em}{}
\titlespacing*{\paragraph}{0pt}{0.4em}{0.5em}

\usepackage{filecontents}

\begin{document}

\date{}

\title{Chameleon: Recovering Cyber-Physical Systems from Memory Corruption Attacks via ML Surrogates}

\author{
{\rm Mohsen Salehi}\\
The University of British Columbia\\
Vancouver, Canada\\
msalehi@ece.ubc.ca
\and
{\rm Karthik Pattabiraman}\\
The University of British Columbia\\
Vancouver, Canada\\
karthikp@ece.ubc.ca
} 

\maketitle

\begin{abstract}
Cyber-physical systems (CPSs)  
can be compromised through memory corruption vulnerabilities, which can result in safety violations. 
Existing techniques mostly focus on detecting such attacks but respond by terminating or
halting execution upon attack detection, which is not acceptable in CPSs
as interrupted tasks can have catastrophic consequences.
Other techniques replace compromised CPS components with simplified defaults 
that degrade system behavior, or reboot the system upon attack detection, which are not suitable for CPS deployed in safety-critical domains. 

We propose \technique{}, a novel framework for automatically recovering CPSs from
memory corruption attacks using machine learning (ML)-based surrogates trained at compartment granularity
that nearly replicate their original compartments' behavior but 
 are implemented differently, and hence are unlikely to have 
 the same memory corruption vulnerabilities. 
Upon attack detection, \technique{} replaces the compromised
compartment with its trained ML surrogate. 
We implemented \technique{} using the LLVM compiler, 
and evaluated its efficiency and effectiveness on seven different robotic vehicles (RVs),
including simulated and real ones.
We found that \technique{} can generate surrogates 
that closely approximate the original compartments (with an average 
R$^2$=0.96), 
successfully recover the system despite real-world memory corruption attacks 
and complete their tasks 
while incurring low performance and memory overheads on real RVs. 

\end{abstract}

\section{Introduction}
\label{sec-intro}

Cyber-physical systems (CPSs) are deployed across safety-critical infrastructures from medical devices to robotic vehicles (RVs). 
Most CPS software, including  RVs' firmware, is developed using
languages such as C and C++, which are prone to memory corruption
vulnerabilities~\cite{MITRErank, szekeres2013sok, kim2023patchverif} (e.g., buffer overflows). 
An attacker who exploits such a vulnerability can 
hijack the control flow, and potentially cause loss of control resulting in 
 catastrophic failure and safety violations.  


Existing software defenses such as control flow integrity (CFI)~\cite{arias2015hafix,abadi2009control}
and data flow integrity (DFI)~\cite{castro2006securing} 
can detect memory corruption attacks, but
they either terminate the process, 
or halt execution upon error detection.
However, this response is insufficient for CPSs, as these systems operate continuously in safety-critical environments. For example,   
abruptly terminating or halting execution of an RV  mid-operation can cause it to crash.
Thus we need techniques for recovering CPSs after an attack has occurred, i.e., resilience.


Prior work on resilience can be divided into three main categories. 
(i) physical recovery, 
(ii) software or hardware redundancy, 
and (iii) rebooting or replacing components with simpler versions. 
All of these techniques face fundamental limitations. 
(1) Physical recovery techniques~\cite{dash2021pid,kong2018cyber,dash2024specguard,choi2020software} are only effective against physical attacks such as GPS spoofing 
and not memory corruption attacks. 
(2) Software or hardware redundancy techniques~\cite{forrest1997building,ron2025galapagos,fei2018cross} are effective against random faults 
but provide only partial protection against memory corruption attacks.
Homogeneous redundancy offers no protection, 
since all instances share the same underlying vulnerabilities.
In contrast, heterogeneous redundancy may survive an attack on one board, 
but it is costly, as it requires diverse hardware and software implementations. 
Further, it is expensive to run multiple instances simultaneously on 
these resource-constrained devices.
(3) 
Rebooting system components~\cite{qin2005rx,rinard2004enhancing,huang1995software}, 
or replacing the compromised system part with a simplified instance, as in the Simplex architecture~\cite{bak2009system}
(e.g., returning a default value~\cite{li2025software}), either 
have high recovery times or exhibit unsafe behavior, making them impractical for real-time and safety-critical CPSs. 
To address this gap, we propose a component replacement technique to recover CPS from memory corruption attacks without violating safety or real-time constraints. 
\textit{Our main insight is to replace the compromised part of the system at the compartment granularity\footnote{We adopt the definition of a compartment from ACES~\cite{clements2018aces} - a compartment consists of an isolated code region together with its associated data, peripherals, and permitted control transfers. Each instruction belongs to exactly one compartment.} 
with a behaviorally equivalent form in a different representation,  
such as an arithmetic form, 
which is unlikely to contain the same memory corruption vulnerability.}

We use machine learning (ML) to learn the behavior of the compartments for replacement. 
This is backed by our observation that many aspects of the CPS perform computations exhibiting consistent behavioral patterns 
that can be learned via ML techniques    
such as neural network (NN) models. 
This insight is inspired by a long line of work in approximate computing~\cite{esmaeilzadeh2012neural,moreau2015snnap} and high-performance computing (HPC)~\cite{schutt2017schnet,sadowski2016synergies,dong2023auto}, 
where parts of an application are replaced with similar ML models for \textit{performance acceleration}. 

To this end, we propose \technique{}, a technique for recovering CPSs from memory corruption attacks. \technique{} replaces the compromised compartment with an equivalent learned form (i.e., surrogate)
that preserves the same functionality and behavior even under attack. 
\technique{} has two main innovations: 
(1) Training ML-based surrogates at \textit{compartment granularity} that achieve fine-grained behavioral approximation, respect CPS resource constraints, and do not contain the memory corruption vulnerabilities of the original compartment,
and (2) automatically identifying the input-output variable sets of each firmware compartment to enable accurate surrogate training, with \emph{no manual effort from developers}.
\textbf{Innovation 1.} Existing ML-based \emph{detection} methods~\cite{junejo2016behaviour,abbaspour2016detection} train the model on the entire system as
a blackbox (i.e., the input-output relationship of the entire CPS), 
limiting their detection to coarse-grained anomalies in overall system behavior.  
These techniques focus on detection, rather than recovery. 
If these techniques are adapted for recovery, 
they would require replacing the entire system even 
when only a single component is compromised, removing features the system needs for mission completion.
On the other hand, training surrogates at the function level improves accuracy but incurs 
high performance overheads given the large number of functions in CPSs' software (a few thousands). 

To address this granularity trade-off, \technique{} uses software compartmentalization~\cite{clements2018aces}, 
which provides a balanced level of granularity 
to partition the CPS firmware, and train an ML-based surrogate for each compartment 
to approximate its behavior while meeting CPS resource constraints.
However, training surrogates at the compartment level introduces the challenge 
of determining the appropriate input-output variable sets to collect and train 
on for closely approximating each compartment's behavior, which \textbf{Innovation 2} addresses.


\textbf{Innovation 2.} 
In contrast to coarse-grained recovery and detection approaches that 
only consider physical inputs and outputs of an entire CPS 
(e.g., sensor measurements in RVs), \technique{} employs automated dataflow analysis 
to identify the input-output variable sets and CPS state variables 
(e.g., position, velocity, sensor measurements) at   the compartment level. 
This enables \technique{} to collect training data  using the identified variable sets and train an ML surrogate 
to approximate each compartment's behavior (in benign environments) 
without inheriting its vulnerabilities.

At runtime, \technique{} leverages existing detection techniques (e.g., CFI and DFI)~\cite{li2025software}
to detect memory corruption attacks and identify the corresponding compromised compartment.
Upon attack detection, \technique{} disables the compromised compartment and replaces it with the trained ML surrogate, 
to compensate for the disabled compartment and maintain the system's functionality.
Since the surrogate is an ML model composed of arithmetic operations, 
it is unlikely to have the same exploitable memory corruption vulnerabilities. 
Therefore, the CPS is able to complete  its tasks with nearly the same functionality even under attack (as we show later).



\textit{To the best of our knowledge, \technique{} is the first attack recovery technique that automatically generates surrogate 
compartments with similar behavior to the original compartments using ML models, 
and replaces the compromised compartment at runtime upon attack detection with the corresponding trained surrogate 
to  successfully recover the system from memory corruption attacks. }
We use RVs for evaluating \technique{}   because they are widely deployed; 
however, \technique{} can be generalized to other CPSs. 

\textbf{\textit{Contributions.}} We make four contributions as follows.

\begin{itemize}
    \item Propose a recovery technique that trains an ML-based surrogate at the compartment level to approximate each compartment's behavior and, upon attack detection, 
    replace the compromised compartment with its trained ML-based surrogate. 
    \item Propose an automated dataflow analysis that 
    identifies input-output variable sets for each compartment and collects training datasets across different tasks. 
    \item Design \technique{}, a framework that integrates the above techniques to enable CPSs to recover from memory corruption attacks. 
    \technique{} is implemented using the LLVM compiler~\cite{llvmcomp}, making it portable across different CPS devices.
    \item Evaluate \technique{} on four simulated and three real RVs, all running one of two widely used open-source autopilot systems, 
    ArduPilot or PX4, across a diverse range of scenarios and missions. 

\end{itemize}

The results demonstrate that (1) \technique{} can generate surrogate 
compartments 
that closely approximate compartment behavior, achieving an average R$^2$=0.96. 
(2) \technique{} successfully completes multiple types of missions in the attack-free case  
achieving a mean deviation of 5.4\,$\pm$\,2.2m from the original destinations.
(3) 
Across five missions, 
under attack scenarios, \technique{} successfully recovers all missions 
with a mean deviation of 7.8m, while Gecko~\cite{li2025software}, an approach
that replaces the compromised compartment by 
returning default values results in a mean deviation of $48.2$m (a deviation $>$\,10\,m is considered a failure), 
(4) \technique{} incurs an average performance overhead of 8.5\%
and an average firmware size overhead of $\approx$9\% on real RVs. 

\section{Background: Robotic Vehicle Control}
\label{sec-back}

\begin{figure}[h]    
    \centering
    \includegraphics[scale=0.75]{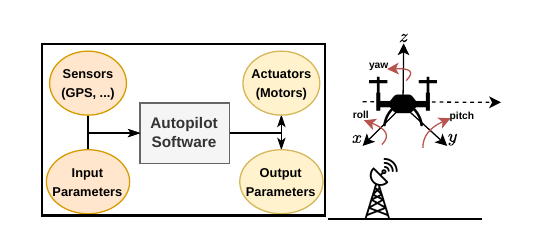}
    \caption{RV Control Loop.}
    \label{subfig:RVLoop}
\end{figure}

As shown in Figure~\ref{subfig:RVLoop}, RVs rely on a combination of autopilot software and hardware components, 
including onboard sensors and actuators such as GPS and gyroscope, to execute their missions. 
ArduPilot~\cite{ardupilot} and PX4~\cite{px4} are two widely used open-source autopilot software platforms
that implement the core control algorithms and system management functions required for autonomous operation.

In addition to flight control, autopilot software supports the following functionalities: 
\textit{ground communication}, which maintains telemetry with a ground control station (GCS); 
\textit{mission management}, which involves defining mission plans and controlling operational modes such as takeoff and landing; 
and \textit{safety monitoring}, which handles obstacle avoidance.
Autopilot software operates as a periodic workload executed in a loop, 
including \textit{tasks} such as $update\_receive$, 
which processes incoming MAVLink messages from the GCS.
Autopilot software runs in both simulated and physical vehicles.
In simulation, environments such as SITL~\cite{ardupilot} and Gazebo~\cite{koenig2004design} reproduce real-world conditions, 
enabling safe testing without the cost and risk of physical deployment. 
In real RVs, the autopilot software is embedded within the firmware, 
where it directly interacts with physical sensors and actuators.

The onboard sensors capture the attitude and movement of an RV across six degrees of freedom (6DoF), 
which include three axes (x, y, z) and three rotational directions: roll, pitch, and yaw. 
These sensor measurements, such as accelerometer and gyroscope readings, alongside other inputs received from the ground control station, 
are fed into the main control loop in the autopilot software. The control loop then computes the appropriate outputs through a controller module, 
such as a proportional-integral-derivative (PID) controller, based on the current vehicle state.
Finally, the resulting signals are forwarded to actuators, such as motors, to drive the vehicle's physical behavior. 

\label{backsec}

\section{Motivation and Threat Model}
\label{sec-motivation}
In this section, we explain the main limitations of current defense techniques (\S\ref{sub-sec:current-limitations}), 
describe how \technique{} addresses these limitations (\S\ref{sub-sec:ourapproach-solution}),  
and discuss the assumptions and threat model of our work (\S\ref{subsec-assumption}). 
Finally, we show an example of how RVs operate and the insight behind \technique{} (\S\ref{subsec:example}).

\subsection{Current Techniques' Limitations}
\label{sub-sec:current-limitations}

\textbf{L1: No Recovery.}
Although many detection techniques have been proposed to address memory corruption attacks, 
they do not provide recovery after attack detection. 
For instance, techniques such as control flow integrity (CFI)~\cite{etigowni2018crystal,burow2017control,xia2012cfimon,davi2015hafix} 
and memory protection~\cite{almakhdhub2020mu,zhou2020silhouette,clements2017protecting}
terminate the process or halt execution upon detection, 
 resulting in task failure.
Other techniques require hardware redesign to prevent or detect memory corruption
vulnerabilities~\cite{ziad2021no,koeberl2014trustlite}. This makes them unsuitable for many CPS deployments and  
they also provide no recovery upon attack detection.

\textbf{L2: Incur High Overheads.} 
Another class of techniques uses redundancy, either in software, 
such as N-version programming~\cite{forrest1997building,avizienis1985n,ron2025galapagos}, or in hardware, such as BlueBox~\cite{fei2018cross}, to recover from attacks. 
These approaches run multiple instances simultaneously and switch to a backup when an attack is detected. 
Other techniques simply reboot the system upon attack detection~\cite{abdi2018guaranteed}.
These techniques incur high overhead or depend on rebooting upon attack detection, which makes them unsuitable for CPS that are often resource-constrained and have real-time requirements.

\textbf{L3: Physical-attacks Only.} 
These techniques~\cite{dash2024specguard,choi2020software} 
replace the PID controller or sensors with ML-trained models.
However, these techniques are  designed for physical attacks on RVs' sensors, 
whereas in memory corruption attacks, the attacker exploits memory vulnerabilities to manipulate software behavior, such as hijacking the control flow. 
Thus, they are 
not sufficient to recover from such attacks. 

\textbf{L4: Removing Functionalities.} 
Some recovery techniques have been proposed to replace a compromised component, 
whether hardware or software, with a simpler version to recover the device from attacks. 
Choi et al. propose a technique for RVs~\cite{choi2020software} that replaces a compromised sensor with a software prediction, under physical attacks. 
In another approach for CPS~\cite{li2025software} called Gecko, 
a compromised software component is replaced 
with a degraded version that returns a default value. 
However, these techniques remove the main functionality,  
which 
can 
cause system failure, thus compromising safety.

\textbf{L5: Incorrect Granularity.} 
Recent techniques have leveraged advances in ML to improve the security of CPSs. 
However, these techniques have two main limitations. 
First, they focus on detection~\cite{ding2021mini} (L1) or recovery from physical attacks in RVs~\cite{dash2021pid} 
(L3). 
Second, most of them~\cite{chen2018learning,goh2017anomaly,junejo2016behaviour} treat the entire system as a blackbox 
and rely only on sensor measurements. 
If these techniques are adapted for recovery, they replace the entire system with a blackbox-trained model that lacks 
the fine-grained features 
often required for mission completion, such as updating CPS state variables.
Mini-Me~\cite{ding2021mini} trains a model for the controller function 
to estimate its outputs, detect discrepancies, and trigger the fail-safe mechanism (land the vehicle); however, 
it is also inapplicable to recovery because 
any part of the firmware may be compromised, requiring a recovery technique capable of replacing arbitrary components and continuing the mission.
Moreover, training a separate ML model for every function in the firmware is impractical.
\subsection{Our Approach: \technique{}}
\label{sub-sec:ourapproach-solution}
We introduce \technique{}, a new recovery technique that provides three solutions (S1-3) 
to address the current techniques' limitations (L1-L5).

\textbf{S1: \textit{Addressing L3 and L5}.}  
\technique{} trains ML-based surrogates at a finer granularity than prior ML-based detection techniques, 
which treat the entire CPS as a blackbox (L5), to approximate the behavior of each compartment, 
enabling replacement of only the compromised component with its surrogate rather than the entire system.
To do so, \technique{} partitions CPS firmware into task-level compartments~\cite{clements2018aces} (detailed in \S\ref{sec-design:compartmentalization}), 
thereby avoiding the need to train and deploy a separate model for every firmware function, which is impractical for CPSs. 
Since ML surrogates are written in a different representation (e.g.,  arithmetic operations), they are
unlikely to contain the same memory-unsafe operations and as a result the same 
memory corruption vulnerabilities (L3). 



\textbf{S2: \textit{Addressing L1, L2, L3, and L4}.}  
At runtime, \technique{} leverages existing CFI and DFI techniques to monitor the compartments and detect memory corruption attacks (L3).
Upon attack detection, the compromised compartment is isolated and replaced with its trained ML surrogate, preserving the compartment's functionality (L4) 
and allowing the CPS to complete its task despite the attack (L1).
Therefore, \technique{} requires neither hardware modifications nor redundancy in running simultaneous instances (L2).

\textbf{S3: \textit{Addressing L5}.} 
To train an ML-based surrogate at a finer granularity (i.e., compartment level), 
we propose an automated dataflow analysis that employs two heuristics (discussed in \S\ref{sec-design:dataflow}) to identify the entry and exit functions of each compartment, the corresponding input and output variables, 
and the relevant CPS state variables (e.g., position and velocity in RVs), 
to capture each compartment's behavior, which is then used to collect the training dataset.

Table~\ref{tab:limiations} summarizes the prior techniques and their limitations. 
\textit{As shown, none of the prior techniques addresses all limitations entirely, while \technique{} does.}

\begin{table}[h]
    \centering
    \caption{Summary of current techniques and their limitations. \halfcirc: some approaches in the category do not have the limitation. "-": the limitation is not applicable.}
    \label{tab:limiations}
    \resizebox{\columnwidth}{!}{%
    \begin{tabular}{c|c:c:c:c:c}
    \textbf{Techniques} & \textbf{\makecell{Recovery?\\ (L1)}} & \textbf{\makecell{Low Overhead\\ (L2)}} & \textbf{\makecell{Attack Types\\ (L3)}} & \textbf{\makecell{Keep Functionality\\ (L4)}} & \textbf{\makecell{Fine-grained\\ (L5)}} \\
    \hline\hline
    Traditional Detection                  & \XSolidBrush    & \halfcirc  & Memory     & -                           & \halfcirc   \\
    N-version Programming                  &  \Checkmark   & \XSolidBrush     & -                     & \Checkmark   & \halfcirc       \\
    Physical Recovery               & \Checkmark & \halfcirc     & Physical      & \halfcirc       & \XSolidBrush       \\
    Gecko                                  & \Checkmark & \Checkmark  & Memory     & \XSolidBrush       & \Checkmark  \\ 
    \hline
    \textbf{\technique{}} & \Checkmark & \Checkmark & Memory     & \Checkmark   & \Checkmark  \\
    \hline
    \end{tabular}
    }
\end{table}


\subsection{Assumptions and Threat Model}
\label{subsec-assumption}

\textbf{Assumptions.}
We make the following four assumptions.
First, we assume that \technique{} has access to either the firmware's source code 
or its LLVM intermediate  representation (IR)~\cite{llvmcomp}, 
since \technique{} operates on LLVM IR.

Second, we assume that an attacker cannot interfere with the training process, 
including poisoning the training data for the ML models.
This is reasonable because NN training is typically
conducted in a controlled environment that is
isolated from external modifications~\cite{amodei2016concrete}.

Third, we assume the existence of a security reference monitor (i.e., CFI and DFI) 
that can detect when a memory corruption vulnerability (e.g., buffer overflows) has been exploited. 
This is a standard assumption in recovery techniques~\cite{li2025software,dash2024specguard} -  
 attack detection is orthogonal to our technique. 

Fourth, 
hardware components (e.g., I/O peripherals), the privileged software components (e.g., firmware), 
and \technique{}'s own components are considered trustworthy.

\noindent\paragraph{\textbf{Threat Model.}}
We consider an adversary who can carry out remote attacks against the target system without having root privileges or physical access.
We further assume that the core system software is written by honest developers and is benign; however, it may include memory corruption vulnerabilities such as buffer overflows.
An adversary who discovers and exploits such a vulnerability 
can redirect the firmware's control flow 
and invoke security-critical functions 
through code-reuse attacks (control-flow violations, \S\ref{Sec:Back-memory}).
Further, we do not consider code injection attacks due to the presence of memory protection unit (MPU)~\cite{clements2018aces,li2025software}.
Hardware-level corruptions, physical attacks such as sensor spoofing, data-only 
attacks, and side-channel attacks are outside the scope of this work.

\subsection{Motivating Example: Robotic Vehicles}
\label{subsec:example}

To illustrate the main insight behind \technique{}, we use an example drawn from RVs. 
Figure~\ref{fig:motivation-example} shows an example function in PX4, a popular autopilot used in RVs, 
containing a known buffer overflow vulnerability~\cite{stackoverflow,kim2018securing} (Line 3).
If an attacker exploits this vulnerability, they can hijack the
control flow of safety-critical components such as the flight control program, potentially causing the RV to crash.
\begin{figure}[ht]
\begin{lstlisting}[style=cppstyle, columns=flexible, basicstyle=\ttfamily\scriptsize, lineskip=-0.5pt]
MultirotorMixer * MultirotorMixer::from_text(\*arguments*/) {
	// Allocation Variables
	if (sscanf(buf, "R: %s %d %d %d %d%n", geomname, &s[0], &s[1], &s[2], &s[3], &used) != 5) {
		debug("multirotor parse failed on '%s'", buf);
		return nullptr;
	}
	// ...
}
\end{lstlisting}
\caption{Real buffer overflow vulnerability in PX4~\cite{stackoverflow}.}
\label{fig:motivation-example}
\end{figure}

The main insight behind \technique{} is that replacing the vulnerable task with an ML-based surrogate, 
such as a long short-term memory (LSTM) network with fixed weight matrices, makes it unlikely 
	for the exploit to succeed, 
since such a surrogate uses no user-controlled buffer lengths, no pointer arithmetic, and no dynamic memory allocation during inference.
While inference libraries are themselves written in C/C++ and may contain memory corruption vulnerabilities,
the attack surface \emph{differs} fundamentally: the original vulnerability is triggered by 
attacker-controlled byte streams with user-controlled lengths,
whereas the surrogate receives only a fixed-size vector of typed integer or floating-point values.
Thus, the original exploit is unlikely to work on the surrogate.

\label{motivsec}

\section{Design: \technique{}}
\label{sec-design}



\begin{figure*}[ht]    
	\centering
	\includegraphics[scale=0.75]{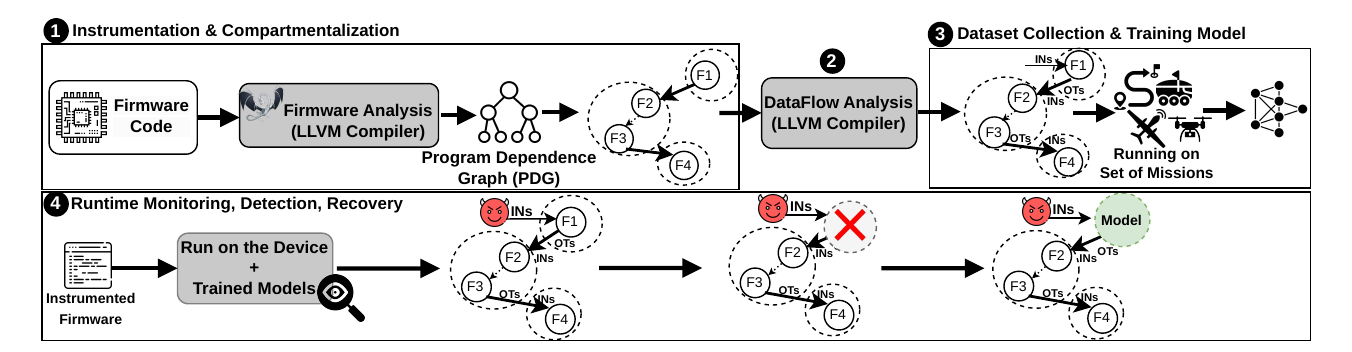} 
	\caption{Overview of \technique{}'s operation, including its four main phases: the first three are performed offline, while the fourth is performed online.}
	\label{img:Overview}
\end{figure*}
In this section, we present the design of \technique{}, our attack recovery technique for CPSs that addresses the limitations discussed in \S\ref{sec-motivation}.
Figure~\ref{img:Overview} shows \technique{}'s workflow, consisting of four main phases (labeled with black circles): (1) compartmentalization and instrumentation, 
(2) dataflow analysis, (3) data collection and training, and (4) runtime monitoring.
\technique{} generates ML-based surrogates that replicate each compartment's behavior 
enabling runtime replacement when a compartment is exploited, thereby allowing the CPS to recover and continue its task.
We describe the different phases of the system  (\S\ref{sec-design:compartmentalization}-\S\ref{sec-design:runtime}).
\textit{The first three phases of \technique{} (i.e., Compartmentalization \& Instrumentation, Dataflow Analysis, and Model Construction) are all executed offline.}

\subsection{Compartmentalization \& Instrumentation}
\label{sec-design:compartmentalization}


\noindent
\textbf{\textit{Compartmentalization.}}
Unlike prior ML-based detection techniques~\cite{chen2018learning,goh2017anomaly,junejo2016behaviour,abbaspour2016detection} that consider the entire system as a blackbox and train models 
on the input-to-output relationship of the entire system, \technique{} trains surrogates at a finer granularity to improve behavioral approximation 
accuracy and replace only the compromised part rather than the entire system (S1 in \S\ref{sub-sec:ourapproach-solution}).

A straw man solution to address this limitation would be training a surrogate for each function inside CPS firmware, but this would incur significant overhead
and is hence not suitable for resource- and time-constrained CPSs. 
Therefore, there is a trade-off among performance, memory overhead, and model accuracy in terms of behavioral similarity to the original firmware part.
\textit{Our key insight is that the optimal granularity is to partition firmware into groups of closely related functions that together represent a CPS task, 
and frequently interact at runtime, minimizing cross-partition calls while keeping each partition semantically meaningful in terms of behavior.}
Thus, \technique{} uses firmware compartmentalization~\cite{clements2018aces,li2025software} 
to partition the firmware based on CPS tasks (e.g., $update\_altitude$ in RVs), 
enabling both efficient surrogate training for each compartment instead of each function
and efficient recovery by replacing only the exploited partition rather than the entire system.

Firmware compartmentalization is formulated as a graph partitioning problem~\cite{clements2018aces}, where the process begins
by constructing a program dependence graph~\cite{ferrante1987program} (PDG) 
that captures all control-flow, global data, and peripheral dependencies of the firmware, with type-based alias analysis~\cite{lattner2007making} used to resolve pointer targets, as done in prior work~\cite{clements2018aces}.
Based on factors such as performance constraints and the compartmentalization policy (i.e., task-based partitioning in \technique{}),
the algorithm groups related functions into individual partitions called compartments, aiming to minimize calls between compartments at runtime.
Additionally, all data shared across compartments, such as global variables, are grouped into a shared compartment. 
At the end of the compartmentalization process, the firmware is partitioned into different compartments
based on tasks. 

\noindent
\textbf{\textit{Instrumentation for Detection and Recovery.}}
After the firmware is divided into different compartments, it needs to be instrumented
for three purposes: (1) monitoring inter-compartment transitions, 
(2) checking CFI and DFI within compartments, 
and (3) replacing the compromised compartment with its trained surrogate upon attack detection.
For (1) and (3), \technique{} instruments compartment transitions
to invoke the \monitor{}, which verifies the validity of each transition according to the PDG.
Also, the instrumentation allows \technique{} to replace the compromised compartment
with its corresponding trained surrogate upon detecting an attack,
compensating for the disabled compartment (\S\ref{sec-design:runtime}).
For (2), \technique{} monitors the control flow of indirect calls and
memory writes within each compartment. It detects an attack when one of them constitutes a malicious write or control transfer, 
and identifies the compromised compartment at runtime.
This part of the process (i.e., (1) monitoring compartment transitions
and (2) CFI/DFI checking) is orthogonal to \technique{}~\cite{clements2018aces,li2025software}.

\subsection{Dataflow Analysis}
\label{sec-design:dataflow}

Once compartmentalization is complete and the compartments are identified, the next phase of the \technique{} process is static dataflow analysis
to identify (1) the entry and exit functions of each compartment 
and (2) the input and output variable sets of each compartment.
The key insight is that, in order to train a surrogate that accurately 
approximates the behavior of the original compartment efficiently 
for resource-constrained CPSs, we train the model based on entry and exit functions, 
since they characterize each compartment's behavior, 
rather than considering all functions within each compartment.
Therefore, these identified entry and exit functions, along with their input and output variable sets, are then used in the next phase
to collect the training dataset and train a surrogate for each compartment (\S\ref{sec-design:model}).

To perform this phase automatically, \technique{} provides a component called
\textit{\identifier{}} that implements two heuristics in two sub-processes. 
First, the \identifier{} takes the compartments and generated PDG as input and outputs the list of
entry and exit functions for each compartment.
To do so, it performs static analysis on the call instructions (in LLVM IR) of each function in a compartment and applies three conditions.
First, if a function in one compartment invokes a function in another compartment, the \identifier{} records the caller as an exit function for its compartment and the callee as an entry function for its corresponding compartment (e.g., F1 and F2 in Figure~\ref{img:Overview}).
Second, if a function has outgoing calls but no other function ever calls it, it is identified as an entry function, since such functions are typically invoked directly by the system scheduler (e.g., the controller in RVs) rather than by other firmware functions.
Third, if a function is called by other functions within the same compartment but itself makes no calls, it is identified as an exit function, as it represents a terminal computation node whose output propagates either upward through the compartment's call chain or to the user. 

Algorithm~\ref{alg_dataflow_functions} demonstrates how the \identifier{} identifies the entry and exit functions for each compartment.
As an initialization step (lines 2--11), \identifier{} first iterates through all functions and their call instructions in each compartment to populate two auxiliary sets: $CalledFuncs$, which records every function that is called by at least one other function (line 9), and $IntraCalledFuncs[C_i]$, which records callees that reside within the same compartment (lines 10--11).
In the next step (lines 12--22), for each call instruction $I$ of every function $f$ in compartment $C_i$, if the callee belongs to a different compartment $C_j$, the \identifier{} records $f$ as an exit function of $C_i$ and the callee as an entry function of $C_j$ (lines 16--18).
After processing all call instructions of $f$, the two additional conditions are checked: if $f$ has no callers anywhere in the firmware, i.e., $f \notin CalledFuncs$, yet has at least one call instruction, i.e., $\text{calls}(f) \neq \emptyset$, it is added to $EntryFuncs[C_i]$ (lines 19--20); if $f$ is called by at least one function within the same compartment, i.e., $f \in IntraCalledFuncs[C_i]$, but has no call instructions, i.e., $\text{calls}(f) = \emptyset$, it is added to $ExitFuncs[C_i]$ (lines 21--22).
The output consists of the $EntryFuncs$ and $ExitFuncs$ sets for each compartment (line 23).

\begin{algorithm}[h]
	\begin{footnotesize}
		\caption{Identifying Entry and Exit Functions}
		\label{alg_dataflow_functions}
		\SetKwInOut{Input}{inputs}
		\SetKwInOut{Output}{output}
		\SetKwProg{identifyEntryExit}{identifyEntryExit}{}{}
		\SetKwFor{ForEach}{foreach}{do}{endfch}
		\identifyEntryExit{$(Compartments)$}{
			\Input{Set of compartments $\mathcal{C} = \{C_1, \dots, C_n\}$, each $C_i$ containing a set of functions; $\text{calls}(f)$ = the set of call instructions in $f$}
			\Output{$EntryFuncs[C_i]$, $ExitFuncs[C_i]$ for each $C_i \in \mathcal{C}$}

			\ForEach{compartment $C_i \in \mathcal{C}$}{
				$EntryFuncs[C_i] \leftarrow \emptyset$\;
				$ExitFuncs[C_i] \leftarrow \emptyset$\;
				$IntraCalledFuncs[C_i] \leftarrow \emptyset$\;
				\ForEach{function $f \in C_i$}{
					\ForEach{call instruction $I$ in $f$}{
						$callee \leftarrow \text{target of } I$\;
						$CalledFuncs \leftarrow CalledFuncs \cup \{callee\}$\;
						\uIf{$callee \in C_i$}{
							$IntraCalledFuncs[C_i] \leftarrow IntraCalledFuncs[C_i] \cup \{callee\}$\;
						}
					}
				}
			}

			\ForEach{compartment $C_i \in \mathcal{C}$}{
				\ForEach{function $f \in C_i$}{
					\ForEach{call instruction $I$ in $f$}{
						$callee \leftarrow \text{target of } I$\;
						\uIf{$\exists\, C_j \in \mathcal{C},\ C_j \neq C_i$ s.t. $callee \in C_j$}{
							$ExitFuncs[C_i] \leftarrow ExitFuncs[C_i] \cup \{f\}$\;
							$EntryFuncs[C_j] \leftarrow EntryFuncs[C_j] \cup \{callee\}$\;
						}
					}
					\uIf{$f \notin CalledFuncs \land \text{calls}(f) \neq \emptyset$}{
						$EntryFuncs[C_i] \leftarrow EntryFuncs[C_i] \cup \{f\}$\;
					}
					\uIf{$f \in IntraCalledFuncs[C_i] \land \text{calls}(f) = \emptyset$}{
						$ExitFuncs[C_i] \leftarrow ExitFuncs[C_i] \cup \{f\}$\;
					}
				}
			}
			\KwRet{$EntryFuncs,\ ExitFuncs$}\;
		}
	\end{footnotesize}
\end{algorithm}

Second, once the entry and exit functions of each compartment are identified, \identifier{} determines the input and output variables of these functions for each compartment.
The purpose of this sub-process is to capture the value changes of these variables along with CPS state variables (e.g., velocity and position in RVs) to construct the input and output vectors of the learning model, which are then used in the next phase, surrogate training (\S\ref{sec-design:model}).

Algorithm~\ref{alg_dataflow_variables} demonstrates how the \identifier{} identifies the input and output variable sets for each compartment.
It takes the list of identified entry and exit functions as input, and iterates through the instructions within all basic blocks of each function (lines 5--14).
Then, it classifies variables according to two criteria: (1) \textit{input} variables are those used but not defined within the function, such as function arguments or global variables (lines 8--10); (2) \textit{output} variables are those assigned a value that is propagated outside the function, excluding purely local variables (lines 11--14).
The \identifier{} repeats this for all entry and exit functions of each compartment and finalizes the $InputVars$ and $OutputVars$ sets (line 15), which, along with CPS state variables, describe the behavior of that compartment and are then used in the next phase for training surrogates (\S\ref{sec-design:model}).

\begin{algorithm}[h]
	\begin{footnotesize}
		\caption{Identifying Input, Output Variables}
		\label{alg_dataflow_variables}
		\SetKwInOut{Input}{inputs}
		\SetKwInOut{Output}{output}
		\SetKwProg{identifyIOVars}{identifyIOVars}{}{}
		\SetKwFor{ForEach}{foreach}{do}{endfch}
		\identifyIOVars{$(EntryFuncs, ExitFuncs)$}{
			\Input{$EntryFuncs[C_i]$, $ExitFuncs[C_i]$ for each compartment $C_i \in \mathcal{C}$}
			\Output{$InputVars[C_i]$, $OutputVars[C_i]$ for each $C_i \in \mathcal{C}$}

			\ForEach{compartment $C_i \in \mathcal{C}$}{
				$InputVars[C_i] \leftarrow \emptyset$\;
				$OutputVars[C_i] \leftarrow \emptyset$\;
				\ForEach{function $f \in EntryFuncs[C_i] \cup ExitFuncs[C_i]$}{
					\ForEach{basic block $B$ in $f$}{
						\ForEach{instruction $I$ in $B$}{
							\ForEach{operand $v$ used by $I$}{
								\uIf{$v$ is not defined within $f$}{
									$InputVars[C_i] \leftarrow InputVars[C_i] \cup \{v\}$\;
								}
							}
							\uIf{$I$ is a store or return instruction}{
								Let $v$ be the value stored or returned\;
								\uIf{$v$ escapes $f$ (non-local)}{
									$OutputVars[C_i] \leftarrow OutputVars[C_i] \cup \{v\}$\;
								}
							}
						}
					}
				}
			}
			\KwRet{$InputVars,\ OutputVars$}\;
		}
	\end{footnotesize}
\end{algorithm}

\subsection{\technique{}'s ML-based Model} 
\label{sec-design:model}
After identifying the task-based compartments and the
input/output variable sets of their entry and exit functions in the prior phases,
this phase proceeds in three steps: (1) instrumenting the firmware to log the identified variable
values along with CPS state variables at runtime, (2) collecting a benign dataset capturing each compartment's
behavior, and (3) training a surrogate ML model for each
compartment offline for runtime deployment (\S\ref{sec-design:runtime}).

\noindent 
\textbf{\textit{(1) Instrumentation for Logging.}} 
To train ML-based surrogates, \technique{} needs to collect a benign dataset capturing each compartment's behavior.
To this end, \technique{} provides an instrumentation component called \logger{}
that takes as input the firmware and the identified entry and exit functions along with
their input and output variables from the previous phase.
\logger{} iterates through the firmware and, upon identifying one of those functions, inserts logging calls at the start and end of that function
to record the value changes of the input and output variables along with CPS state variables at runtime.

\noindent
\textbf{\textit{(2) Dataset Collection.}}
Since there is no standard and public training dataset available for CPSs, 
we need to collect a comprehensive dataset for model training that captures each compartment's behavior from both simulated and real systems
in different scenarios and environmental conditions. 
To do so, we provide an automated component that generates diverse tasks across different scenarios
and executes the instrumented CPS firmware from the previous step.
Then, \logger{} collects the time-series dataset of input/output variable changes
along with the CPS state variables (e.g., velocity, position, body angular rates in RVs) during those missions.
For instance, in RVs (ArduCopter, a firmware variant of ArduPilot~\cite{ardupilot}),
a mission includes taking off, following different paths to different waypoints,
and finally returning to the launch point and landing.
Given the default logging frequency of the system (e.g., 10 Hz in ArduCopter, i.e., every 100 milliseconds),
\logger{} collects 600 data samples (e.g., input and output vector pairs) in a 60-second mission, which are then used for training. 

To train ML-based surrogates that capture the behavior of each compartment, 
\technique{} requires a dataset consisting of sequences of past identified inputs 
along with CPS state variables for that compartment used to predict the corresponding outputs and CPS state variables.
However, the dataset collected through logging is time-series data in a vector-to-vector format.
To improve the accuracy and robustness of \technique{}'s learning model, we apply a sliding window to convert it into a windowed sequential dataset, where each sample pairs a sequence of past observations as input with a single predicted output at the next time step (window-based framing). 
For instance, \technique{} trains surrogate $A$ for compartment $A$ using 
a sequence of past inputs and CPS state variables of compartment $A$ 
within a window of $W$ time steps (e.g., $t_k, \dots, t_{k+W}$) to predict the corresponding outputs and CPS state variables at the next time step $t_{k+W+1}$.
Finally, the output of this step is the processed training dataset, which is used to train the surrogates in the next step.


\noindent
\textbf{\textit{(3) Offline Training.}}
After collecting and processing the dataset, \technique{} uses long short-term memory (LSTM) neural networks,
a type of recurrent neural network (RNN), 
to train a surrogate that approximates the behavior of the corresponding compartment by learning the mapping between its input and output sequences.
\technique{} uses LSTM for the following three reasons:
(1) ML techniques such as logistic regression and random forests, 
are primarily designed for classification or regression on static, 
labeled datasets, whereas our goal is prediction based 
on the sequential history of time-series data.
Furthermore, LSTMs can model nonlinearities, enabling them to learn complex relationships between the inputs and outputs of CPSs~\cite{dash2021pid}.
(2) LSTM uses a sequence of data for training instead of a single input, which improves model accuracy, 
(3) LSTM's gated memory architecture enables the model to retain
and selectively propagate information across long input sequences,
which is critical for capturing the temporal dependencies 
in CPS dynamics, e.g., Euler angles in RVs. 
Convolutional Neural Networks (CNNs) capture only a fixed receptive field and simple RNNs suffer from vanishing gradients,
limiting both for the long-horizon temporal dependencies in CPS dynamics~\cite{bengio1994learning,ding2021mini}.
Our generated LSTM surrogates must be lightweight due to 
resource-constrained CPSs, and also closely approximate the behavior of their compartments. 
To address this, \technique{} employs a 
hyperparameter optimization process to tune key settings such as 
the number of layers, hidden units, and recurrent cell types, 
using cross-validation on different mission datasets to ensure 
accuracy and convergence of the trained surrogate. 
The details are discussed in \S\ref{subsec:evaluation-effectiveness}.
Since the purpose of \technique{} is not classification, we do not use a \textit{softmax} layer.
Furthermore, the model uses 
dropout regularization during training and \textit{ReLU} activation on 
the dense layers, with a final linear output layer for continuous 
regression. The model is trained using mean squared error (\textit{MSE}) 
as the loss function and the \textit{Adam} optimizer.


\subsection{\technique{} Runtime}
\label{sec-design:runtime}
After generating surrogates for each compartment, 
the final phase of the \technique{} process is runtime monitoring: 
monitoring the system and, upon attack detection, 
replacing the compromised compartment with its trained surrogate 
so that the system continues its mission with nearly the same functionality of the replaced compartment. 

This phase has two steps: (1) Attack Detection and (2) Runtime Recovery. 

\noindent
\textbf{\textit{(1) Attack Detection.}}
The instrumented firmware from phase 1 (\S\ref{sec-design:compartmentalization})
is deployed on the system, which adds the \monitor{}
that monitors the inter-compartment transitions, CFI and DFI within
compartments, and detects memory corruption attacks. 
It also identifies the compromised compartment upon detecting an attack.
The \monitor{} then triggers the recovery mode, 
isolates the compromised compartment
and replaces it with the corresponding ML surrogate. 

 \noindent
\textbf{\textit{(2) Runtime Recovery.}}
When the recovery mode is triggered and the compromised compartment is invoked by other compartments or by the system scheduler (e.g., the controller in RVs),
the \monitor{} sends the current input variables and CPS state variables to the corresponding ML surrogate that replaced it.
The surrogate's predicted outputs are then used as the outputs of the compromised compartment instead of the compartment's outputs.

By replacing the compromised compartment with its ML surrogate, \technique{} preserves the compartment's functionality through behavioral approximation. 
Since the surrogate operates in a numerical representation rather than unsafe low-level code, it is unlikely to contain the same memory corruption vulnerability as the original compartment, preventing the attacker from hijacking the control flow after the replacement. 

The surrogates are all stored on-board the device's secondary storage, and upon attack detection, the corresponding surrogate is swapped in to the device's memory. We assume that the device has sufficient secondary storage to store all the surrogates, 
which is the case for RVs. 


\label{designsec}

\section{\technique{} Implementation}
\label{sec:implementation}
\noindent
\textbf{\textit{Compartmentalization \& Instrumentation.}}
\technique{} builds on top of existing compartmentalization and instrumentation
infrastructure~\cite{clements2018aces,li2025software} to partition
the firmware into task-based compartments and instrument it, but modifies the infrastructure to 
invoke \monitor{} at compartment transitions and replace the compromised compartments with their corresponding surrogates upon attack detection.
Both components are implemented as LLVM passes~\cite{llvmpass}. 
\noindent
\textbf{\textit{Dataflow Analysis.}}
This phase is implemented as a component called \identifier{} with two LLVM analysis passes and a Python program that operate on the compartmented firmware's IR.
The first LLVM pass and the Python program identify entry and exit functions 
by iterating over the PDG and call instructions within each firmware's function
and checking the conditions discussed in \S\ref{sec-design:dataflow}.
The second pass takes the identified entry and exit functions of each compartment as input, iterates over the instructions of those functions within their basic blocks, and identifies their input and output variables based on Algorithm~\ref{alg_dataflow_variables}.  

\noindent
\textbf{\textit{Training.}}
\logger{} is implemented as an LLVM module pass that takes the firmware, compartment information, and the identified entry and exit functions of each compartment along with their input and output variables as input.
After collecting data from the instrumented CPS at runtime, an LSTM surrogate model for each compartment is trained offline using TensorFlow 2.18~\cite{abadi2015tensorflow} and Keras 3.10.
Since \technique{} is intended to support resource-constrained devices, including our subject RVs (\S\ref{sub-sec:Experimental}), 
we use 8-bit weight quantization, a well-established neural network compression technique,
to reduce the memory consumption of ML surrogates.

%
\noindent
\textbf{\textit{Runtime Monitoring.}}
The \monitor{} is implemented in C++ and integrated into the instrumented firmware image (e.g., the RV's autopilot software).


\label{implementsec}

\section{\technique{} Evaluation}
This section describes our experimental setup and the evaluation of 
\technique{} across two aspects: effectiveness and efficiency. 
Effectiveness measures \technique{}'s ability to generate surrogates that can approximate compartment behavior and recover the device upon attack detection, 
while efficiency captures the runtime and memory overhead of \technique{}, both during offline training and at runtime.

\subsection{Experimental Setup}
\label{sub-sec:Experimental}
\begin{figure}[ht]
    \centering
    \begin{subfigure}[b]{0.32\columnwidth}
        \centering
        \includegraphics[width=\linewidth,height=1.5cm,keepaspectratio]{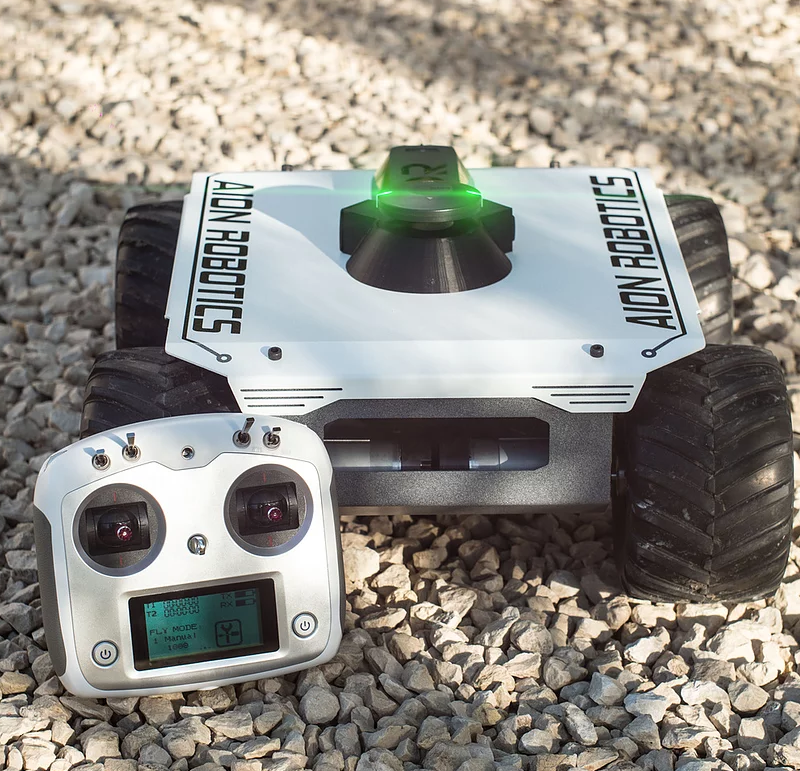}
        \caption{Aion rover}
        \label{fig:aionrover}
    \end{subfigure}
    \begin{subfigure}[b]{0.32\columnwidth}
        \centering
        \includegraphics[width=\linewidth,height=1.5cm,keepaspectratio]{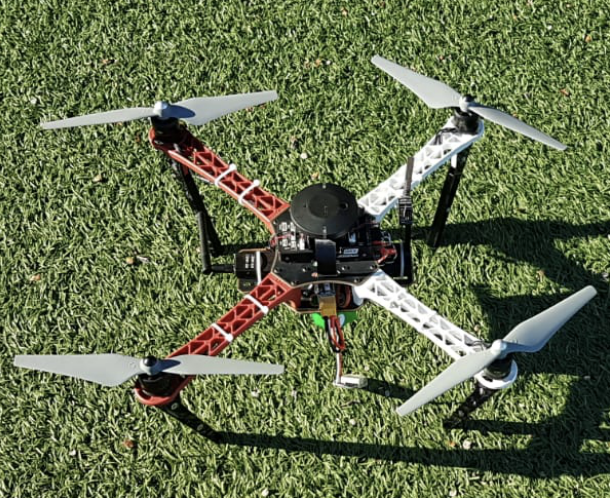}
        \caption{Pixhawk drone}
        \label{fig:pixhawkdrone}
    \end{subfigure}
    \hfill
    \begin{subfigure}[b]{0.32\columnwidth}
        \centering
        \includegraphics[width=\linewidth,height=1.5cm,keepaspectratio]{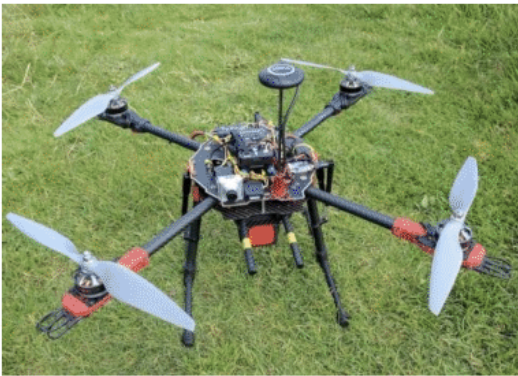}
        \caption{Tarot drone}
        \label{fig:tarotdrone}
    \end{subfigure}
    \caption{Real RV systems used in the experimental setup.}
    \label{fig:rvplatforms}
\end{figure}

\noindent
\textbf{Subject RVs.} We evaluated \technique{} on seven RVs: three real RVs (shown in Figure~\ref{fig:rvplatforms}), an Aion R1 ground rover (Aion rover)~\cite{aionr1},
a DIY drone (Pixhawk drone)~\cite{meier2011pixhawk}, and a Tarot 650 drone (Tarot drone),
and four simulated RV systems,
running across two popular open-source autopilot platforms, ArduPilot~\cite{ardupilot} and PX4~\cite{px4}, specifically:
(1) ArduPilot's quadplane~\cite{ardupilotplane} (ArduPlane),
(2) ArduPilot's quadcopter~\cite{ardupilotcopter} (ArduCopter),
(3) ArduPilot's ground rover~\cite{ardupilotrover} (ArduRover), and
(4) PX4's quadcopter~\cite{px4quadrotor} (PXCopter).
For the simulated systems, we used QGroundControl~\cite{qgroundcontrol}, APM SITL (Software-In-The-Loop)~\cite{ardupilot}, and Gazebo~\cite{gazebo}.
The real RVs are commodity systems based on the Pixhawk platform~\cite{pixhawk}, which is built on an ARM Cortex architecture and integrates ArduPilot or PX4 firmware, a flight management unit (FMU), memory, sensors, and I/O interfaces.

\noindent
\textbf{Compartments.} 
As discussed in \S\ref{sec-design:compartmentalization} and \S\ref{sec-design:dataflow}, 
\technique{} generates task-based compartments for each firmware, and then uses 
\identifier{} to identify the entry and exit functions of each compartment, along with their input and output variable sets. 
These sets are then used to collect data and train a surrogate for each compartment. 
Table~\ref{tab:detailed_results} summarizes the compartments and their associated information across different firmware 
for both simulated and real RVs running ArduPilot or PX4. 
We find that the number of compartments ranges from 12 to 26, with each compartment having an average of 43-96 entry functions and 6-11 exit functions.
The number of input and output variables ranges from 7 to 36 depending on the compartment. 
Finally, the average size of a compartment's surrogate ranges from 211 KB to 221 KB.
%
Our real RVs have 8 GB of secondary storage, along with an average of 2 MB of Flash and 1 MB of RAM as on-board memory. The total size of all the surrogates was at most 5.5 MB, and so they fit comfortably on the RVs' secondary storage. 

\begin{table*}[ht]
    \centering
    \caption{The details of \technique{}'s results across different firmware. 
    The last three columns report the average number of inputs and outputs across the entry and exit functions of all compartments, as well as the average surrogate size
    for each corresponding firmware. Note that the real RVs share the same firmware model as the simulated RVs.}
    \label{tab:detailed_results}
    \resizebox{\textwidth}{!}{%
    \begin{tabular}{c|c|c|c|c|c|c}
    \toprule
    \textbf{Firmware} & \textbf{Number of Compartments} & \textbf{Average Number of Entry Funcs.} &  \textbf{Average Number of Exit Funcs.}  & \textbf{Average Number of Inputs} & \textbf{Average Number of Outputs} & \textbf{Average Surrogate Size}\\
    \midrule
    \makecell{ArduCopter\\(including Pixhawk, Tarot drones)} & 26 & 44 & 6 & 17 & 7 & 211.8 KB\\
    \hline
    ArduPlane & 16 & 43 & 10 & 36 & 14 & 216.0 KB\\
    \hline
    \makecell{ArduRover\\(including Aion Rover)} & 12 & 96 & 8 &  25 & 12 & 217.7 KB\\
    \hline
    PX4 & 14 & 52 & 11 & 28 & 18 & 221.6 KB\\
    \bottomrule
    \end{tabular}
    }
\end{table*}

\noindent
\textbf{Dataset.}
Since no standard publicly available dataset exists for training and evaluating 
\technique{} surrogate models, we conduct more than 100 missions 
for each subject RV with varying mission distances, durations, environmental conditions, 
and flight paths (circular, polygonal, and straight-line paths with multiple waypoints) 
to capture representative real-world RV behavior.
Each mission lasts between 1 and 10 minutes, producing 600 to 6000 timesteps logged at 10 Hz (every 100 ms). 
The collected data (i.e., 100 missions for each RV) is 
then partitioned into 70\% for training, 15\% for validation, and 15\% for testing.




\subsection{Effectiveness}
\label{subsec:evaluation-effectiveness}

\noindent
\textbf{Ablation Studies.}
To select the surrogate model architecture, we evaluate different
candidate designs varying in the number of layers, hidden units,
and recurrent cell types, balancing predictive performance against
model complexity across the subject RVs. We choose an architecture using two
LSTM layers of width 128 each, followed by two fully-connected
layers of 64 and 32 units (230K parameters), as it achieves the best predictive performance-overhead trade-off (R$^2$ = 0.974) 
while requiring fewer parameters than the alternatives. Detailed ablation results are discussed in Appendix~\ref{appendix_sec:ablation}.

\begin{figure}[ht]
    \centering
    \includegraphics[width=0.72\linewidth]{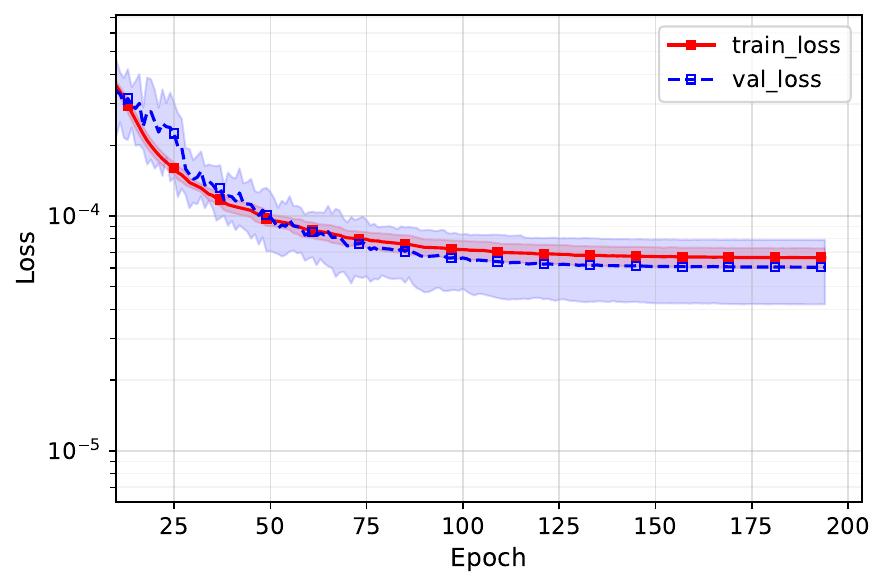}
    \caption{Training and validation loss convergence over 10 independent 
    runs with different random weight initializations. Solid and dashed lines represent the mean training and 
    validation MSE loss, respectively, with shaded regions denoting 
    $\pm$1 standard deviation across runs.}
    \label{fig:training-loss}
\end{figure}

\noindent
\textbf{Surrogate Convergence.}
After identifying the best surrogate architecture, we train it
independently for compartments across the subject RVs. 
To ensure the reliability of the reported results, we conduct 
10 independent training runs 
with different random weight 
initializations. Figure~\ref{fig:training-loss} reports the 
mean training MSE loss (red solid line) and mean validation MSE loss 
(blue dashed line) across all 10 runs at each epoch, where 
the shaded region represents the standard deviation (std-dev) across runs.

We observe that both losses decrease steadily from an initial value of 
approximately $3\times10^{-4}$ and converge within 100 epochs 
to a final value of approximately $6\times10^{-5}$.
The training and validation loss curves remain close to each 
other throughout the entire training process, indicating that 
the model learns generalizable patterns from the training data. 
The narrow standard deviation bands confirm that the selected architecture reaches 
consistent convergence quality across all 10 runs, regardless 
of weight initialization, with a final training loss of $6.64\times10^{-5}$ (std-dev: $6.45\times10^{-6})$ and validation loss of $6.05\times10^{-5}$ (std-dev: $1.85\times10^{-5}$) at the final epoch.
The validation loss remains marginally below the training loss 
throughout training, consistent with the expected behavior of 
regularized neural network models~\cite{srivastava2014dropout}.

\noindent
\textbf{Surrogate Fidelity.}
We evaluate the fidelity of the trained surrogates 
through three aspects: 
(1) average predictive performance measured by R$^2$ and mean absolute error (MAE), (2) time-series prediction on compartment outputs, and (3) the 
deviation in missions' trajectories after replacing a 
compartment with its trained surrogate at runtime.

\textbf{(1) R$^2$ and MAE.} 
We measure the average R$^2$ and MAE of the trained surrogates 
across different compartments of subject RVs discussed 
in~\S\ref{sub-sec:Experimental}.
Across 10 independent training runs, surrogates 
achieve an average of R$^2$=0.96 and MAE=0.20, demonstrating consistent and 
reliable approximation of compartment behavior.

\begin{figure}[ht]
    \centering

    \begin{subfigure}{\linewidth}
        \centering
        \includegraphics[width=\linewidth]{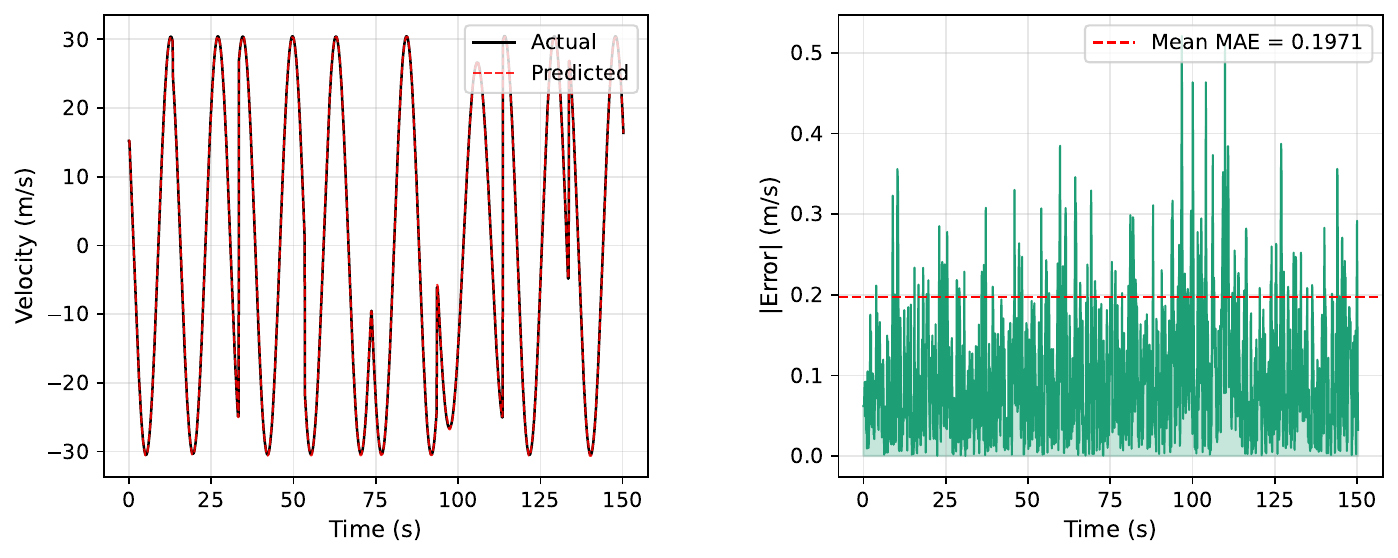}
        \caption{Surrogate model prediction vs. actual Velocity,
                 showing $R^2 = 0.9995$ with a mean absolute error of
                 $0.1971$\,m/s.}
        \label{fig:eval:vn}
    \end{subfigure}


    \begin{subfigure}{\linewidth}
        \centering
        \includegraphics[width=\linewidth]{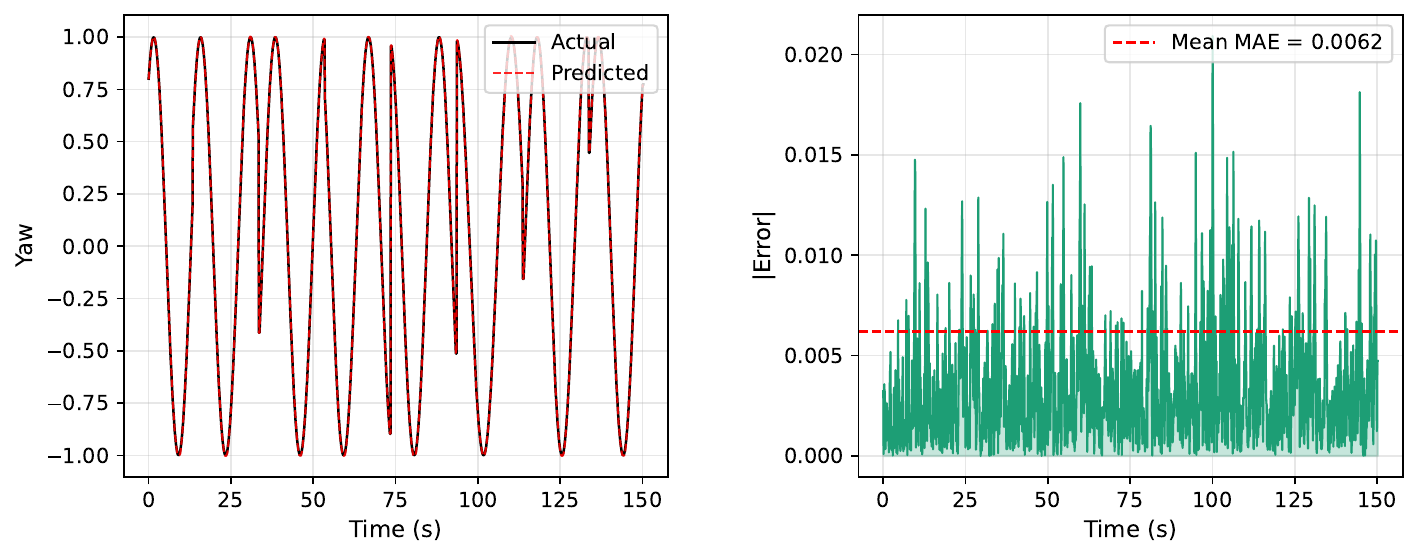}
        \caption{Surrogate model prediction vs. actual Yaw,
                 showing $R^2 = 0.9997$ with a mean absolute error of
                 $0.0062$.}
        \label{fig:eval:gx}
    \end{subfigure}

    \caption{Surrogate model evaluation on two RV state variables.
             Each subfigure shows the predicted vs.\ actual time-series (left)
             and the corresponding absolute error over time (right). Lower values are better.}
    \label{fig:eval:surrogate}
\end{figure}

\textbf{(2) Time-series Prediction.} We illustrate the surrogate 
predictions against the actual value for two of the RV state variables over a 
150-second mission in Figure~\ref{fig:eval:surrogate}. 
For Velocity, the surrogate 
achieves R$^2$=0.9995 with a mean absolute error of 
0.1971~m/s, with the predicted values closely tracking 
the actual measurements across the full dynamic range of 
the signal ($\pm$30~m/s). For Yaw, the surrogate achieves 
R$^2$=0.9997 with a mean absolute error of 0.0062, 
demonstrating precise heading prediction throughout the 
flight. In both cases, the absolute error is small 
and stable over time, confirming that the surrogate 
closely approximates the compartment behavior.

Across the 10 training runs described in \textit{"Surrogate Convergence"},
the surrogate consistently achieves R$^2$=0.9995\,$\pm$\,0.0001
with MAE of 0.197\,$\pm$\,0.006\,m/s for Velocity and 
MAE of 0.006\,$\pm$\,0.0003 for Yaw, demonstrating robust and reproducible approximation of compartment behavior across diverse initializations.

\textbf{(3) Missions Deviation.} To evaluate whether  
surrogates closely approximate their compartments' behavior, we replace different compartments with surrogates 
and run different missions and calculate the deviation from the target destination.
We consider \technique{} as  
successful in recovery 
if at the end of the mission, the deviation from 
the original destination is \textit{less than 10m}.
This threshold accounts for the inherent positioning uncertainty of commodity GPS modules, 
which carry a typical accuracy offset of 5m~\cite{dash2021pid,vadduri2023precise}. Since both the vehicle's reported position and the destination 
coordinates are each subject to this uncertainty, 
these compound to a combined bound of 10m
~\cite{dash2021pid,dash2024specguard}.

\begin{figure}[ht]
    \centering
    \includegraphics[width=0.71\linewidth]{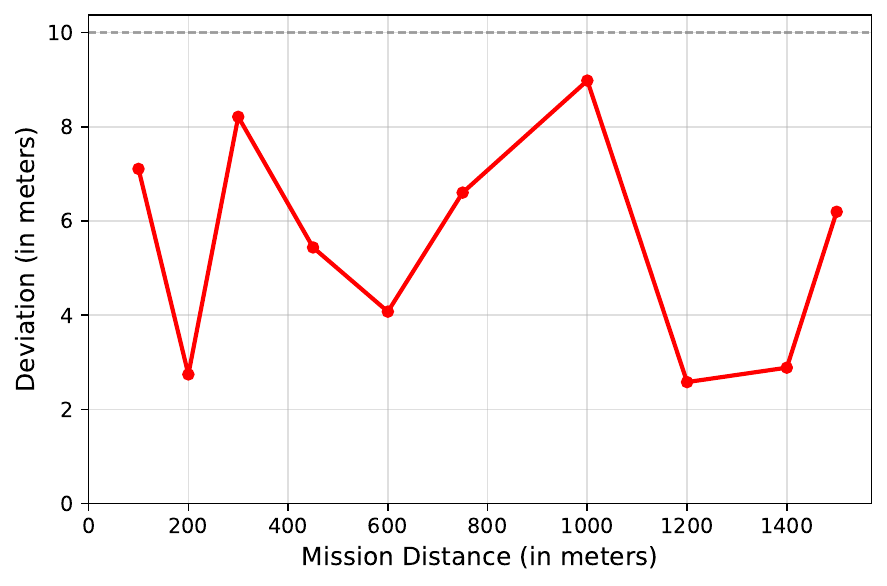}
    \caption{Average deviation from the original destination after replacing the compartment with its surrogate across 10 different missions by \technique{}. Deviation less than 10m indicates mission success.}
    \label{fig:deviation}
\end{figure}

We run 10 different missions on simulated RVs, 
as simulation allows accurate deviation measurement and closely reflects real-system behavior, 
with different scenarios including different distances (100m-1500m) and replace 
the compartment with its trained surrogate and measure the deviation from the original destination.
The results are shown in Figure~\ref{fig:deviation}. 
We find that all 10 missions successfully finish without crashing, and with a mean 
deviation of 5.4\,$\pm$\,2.2m and a maximum deviation of 8.9m, demonstrating that the surrogate model keeps the deviation 
within acceptable bounds (10m). 

\noindent
\textbf{Surrogate Generalization.}
To evaluate generalization across varying mission conditions, 
we assess the surrogate from Figure~\ref{fig:eval:surrogate} 
on different missions. 
Among these, we report results on one of the most challenging missions  (the worst-performing mission in our experiments) - 
a 150-second mission spanning a cruise-to-maneuver transition 
in which the vehicle accelerates from steady cruise into full 
dynamic maneuver.

\begin{figure}[ht]
    \centering

    \begin{subfigure}{\linewidth}
        \centering
        \includegraphics[width=\linewidth]{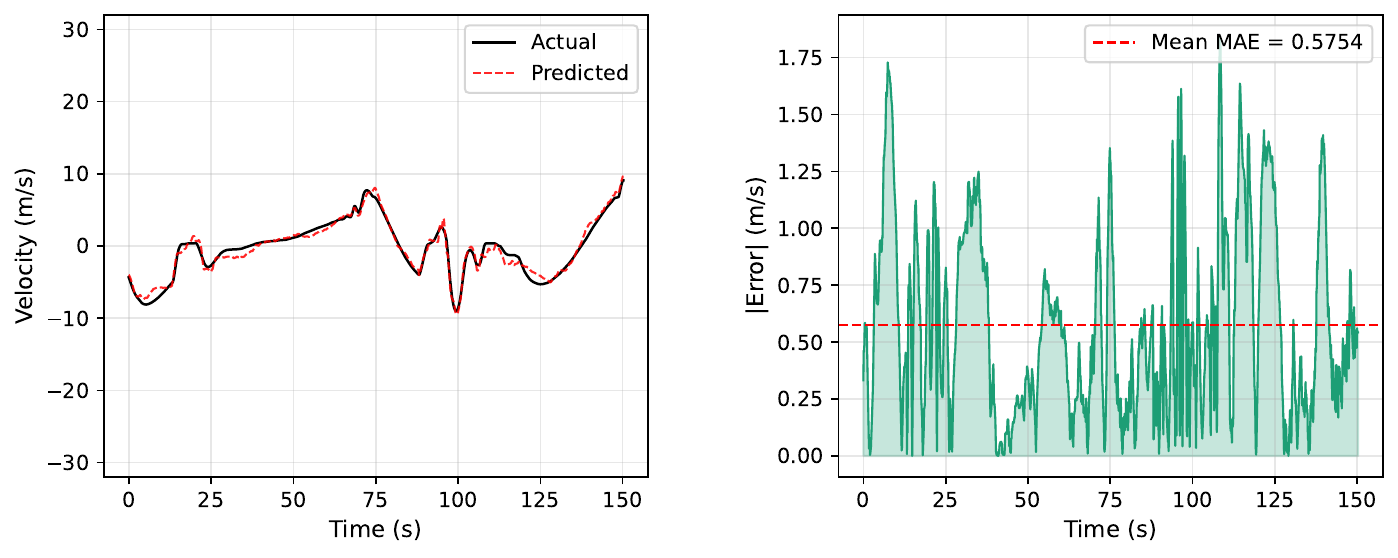}
        \caption{Surrogate model prediction vs. actual Velocity.} 
        \label{fig:eval:vn2}
    \end{subfigure}


    \begin{subfigure}{\linewidth}
        \centering
        \includegraphics[width=\linewidth]{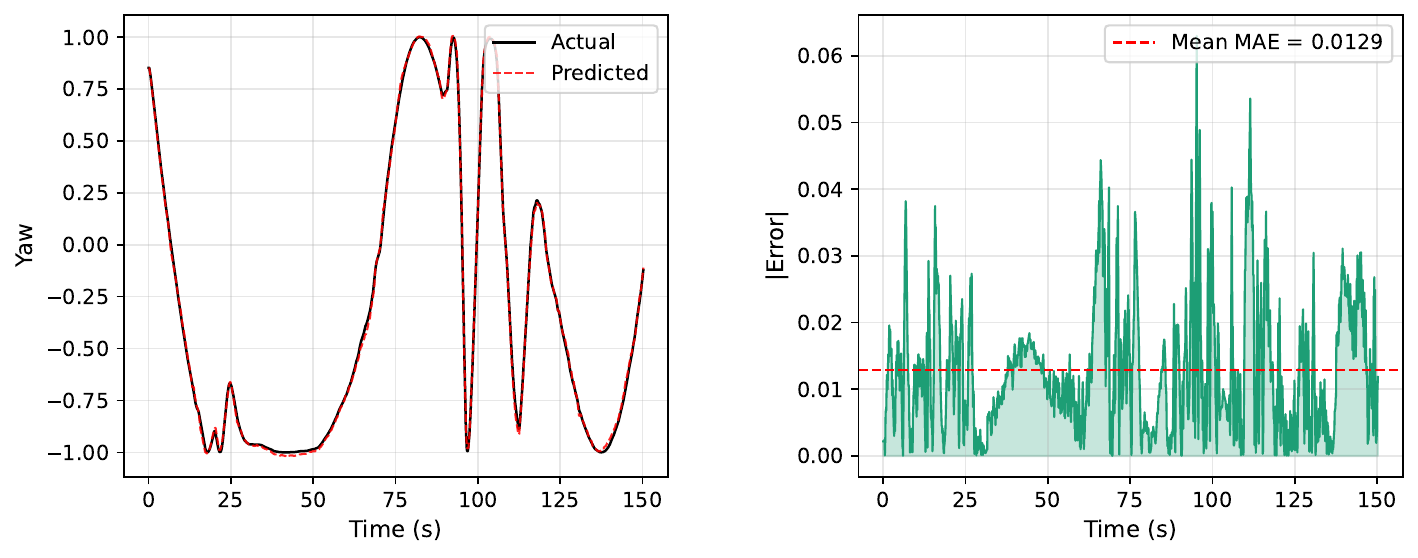}
        \caption{Surrogate model prediction vs. actual Yaw.} 
        \label{fig:eval:gx2}
    \end{subfigure}

    \caption{Surrogate model evaluation on a different complex mission (cruise-to-maneuver transition), 
    demonstrating generalization across varying mission conditions.}
    \label{fig:eval:surrogate2}
\end{figure}

Figure~\ref{fig:eval:vn2} and 
Figure~\ref{fig:eval:gx2} show the surrogate 
predictions against actual values for Velocity and Yaw, 
respectively, with the absolute prediction error shown on 
the right side of each subfigure. 
The surrogate achieves R$^2$ scores of 0.9653 and 0.9994 with mean absolute errors of 0.5754 m/s and 0.0129 for Velocity and Yaw, respectively,
confirming that the surrogate maintains a reasonable  approximation of compartment behavior across different missions without retraining. 

\begin{figure}[ht]
    \centering
    \includegraphics[width=0.72\linewidth]{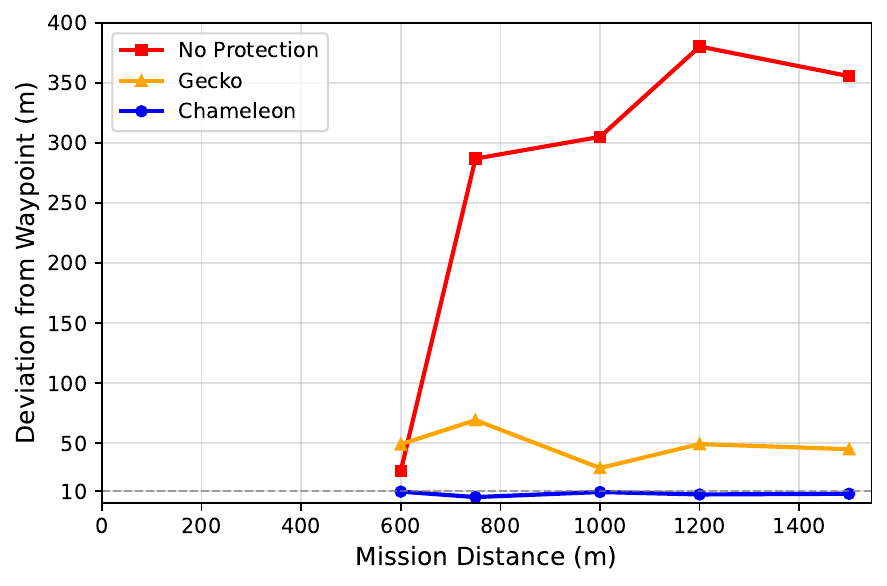}
    \caption{Deviation under attack across five different missions: No-Protection, Gecko, and \technique{}.}
    \label{fig:attack-deviation}
\end{figure}

\noindent
\textbf{Attack Evaluation.}
To evaluate \technique{} in realistic attack scenarios, 
we use the memory corruption vulnerabilities in RVs from prior work~\cite{salehi2026rvdebloater}, 
which were inspired by real-world
vulnerabilities~\cite{stackoverflow,bufferoverflow}. 
A buffer overflow is injected into a function that
processes user input, allowing an attacker to hijack the
control flow of the firmware.
We run five missions with different distances, considering an attacker 
who exploited this vulnerability and injected corrupted data 
to manipulate the mission.
For each mission, we measure the average deviation from the target destination across five runs, 
once \technique{} is triggered and replaces the compromised compartment with its trained surrogate.
As mentioned earlier, a mission is considered successfully recovered if the deviation remains below 10m.
Figure~\ref{fig:attack-deviation} (blue line) shows that all of 
five missions are successfully recovered by \technique{} with a mean 
deviation of 7.8m (range: 5.1--9.6m).
For safety reasons, we performed this evaluation on simulated RVs; 
however, we also tested the feasibility of the attack on the Aion rover,
where an attacker could hijack the control flow and perform malicious actions, such as disarming the motors~\cite{salehi2026rvdebloater}. 

\textit{Comparison with Gecko.}
As mentioned, Gecko~\cite{li2025software} attempts to recover CPS by replacing the compartments with either their default values in an automated manner, or with custom manually-written variants. 
Because the publicly available version of Gecko implements only default-value replacement, we used that design to ensure a fair comparison with \technique{}, which is a fully automated framework\footnote{Gecko's publicly released implementation does not fully realize the system described in its paper (e.g., compartmentalization and recovery upon detection are not fully implemented); 
We therefore re-implemented and fixed the relevant components based on the paper's description with our best effort.}.
%
To evaluate the effects of not preserving the compartments' behavior during recovery, we repeat the  five missions under the same attack, but replacing the 
compromised compartment with its default value as done by Gecko and measure the  deviations.  

As can be seen in Figure~\ref{fig:attack-deviation}, 
Gecko experiences deviations ranging from 29.2m to 69.3m with a mean deviation of 48.2m, which is 6$\times$ the mean deviation of \technique. Thus, with Gecko, all the five missions experienced more than 10m deviations, and hence failed. \emph{This confirms that compartment replacement without behavioral approximation is insufficient for successful recovery.} 
We also observed that during Gecko's recovery, the RV sometimes unexpectedly 
gained altitude before descending and landing or stalled (i.e., froze in place). 
This is because the default value (e.g., velocity) used  
by Gecko includes a vertical component recorded at mission start, 
which no longer reflects the vehicle's actual state at recovery. 

We also evaluated the same missions with no recovery mechanism (i.e., no-protection) under the same attacks.
Without recovery, the RV continues flying with corrupted values
indefinitely; in practice, it would eventually crash (e.g., due to
collisions or battery exhaustion from flying off-course).
Therefore, we force-land it after a time equal to the average of 
\technique{}'s and Gecko's completion time for each corresponding mission 
to ensure a fair comparison.
As shown in Figure~\ref{fig:attack-deviation}, deviations range from 26.9m to 380.3m with 
a mean of 271m, and no mission completes successfully. 
The relatively low deviation in mission 1 (26.9m) is coincidental, 
as the forced landing happened close to the waypoint.  

\textit{Thus, \technique{} consistently recovers all missions successfully 
achieving a 6$\times$ reduction in mean deviation 
compared to Gecko and 34$\times$ compared to no-protection.  }


\subsection{Efficiency}
\noindent\textbf{CPU and Memory Overheads.}
As discussed in \S\ref{sec-design:runtime}, the only phase of \technique{}
that is executed on the system at runtime is \technique{} Runtime.
We measure its CPU overhead on the three real RVs described in \S\ref{sub-sec:Experimental},
as simulated RVs depend on the computing platform and are thus unsuitable for this evaluation.
There is a scheduler in these systems that tracks the total CPU time 
incurred by each task, and we use it to calculate the CPU overhead of \technique{} by analyzing 
the additional CPU time recorded by the scheduler when the RV is equipped with \technique{}. 
We find that the average CPU overhead incurred by \technique{} on real RVs is 8.5\%, ranging from 8\% to 9\%. 
Further, \technique{} does not increase the RV's overall mission time, 
and the RVs satisfy the real-time constraints despite \technique{}.
Finally, the integration of \technique{} into the RV autopilot introduces an average firmware size overhead of $\approx$9\% 
(measured as the percentage of the total on-board memory 
when one surrogate is active). 


\label{evaluatesec}

\section{Discussion}
\label{sec:discussion}


\noindent
\textbf{\textit{Applicability to other CPS.}}
Although we evaluated \technique{} on RVs, \technique{}
does not rely on any RV-specific hardware or firmware.
Since \technique{} is implemented using the LLVM compiler,
it can be applied to other CPSs as long as 
they meet the compartmentalization requirements of ACES~\cite{clements2018aces}, 
namely the presence of an MPU.
Further, the CPS must have sufficient on-board resources 
to store and execute surrogates; this was the case for
the RVs we used. 

\noindent
\textbf{\textit{Detection Techniques.}}
\technique{} relies on existing CFI and DFI techniques for attack detection, whose accuracy affects recovery.
In the case of a false positive, where a benign execution is incorrectly flagged as an attack, \technique{} replaces the compartment with its trained surrogate, which closely approximates the original behavior (R$^2$=0.96 on average, \S\ref{subsec:evaluation-effectiveness}).
Since the surrogate closely preserves the compartment's functionality, a false positive results in only a minor behavioral deviation and is unlikely to disrupt the mission.
Figure~\ref{fig:deviation} further supports this observation, 
where \technique{} replaces the compartment with its surrogate despite no attack occurring. 
Across 10 missions, all missions completed successfully, with a mean deviation of 5.4\,$\pm$\,2.2m and a maximum deviation of 8.9m,
all within the 10m threshold. Thus, a false positive slightly degrades mission precision but does not cause mission failure.
In the case of a false negative, however, where an actual attack goes undetected, the compromised compartment may continue to execute without replacement.
This is a limitation of the underlying detection technique rather than \technique{}; improving the sensitivity of CFI/DFI detection is an orthogonal research direction that, if addressed, would directly benefit \technique{}.

\noindent
\textbf{\textit{Adaptive Attacks.}}
An attacker aware of \technique{} may attempt to subvert the recovery mechanism
through four strategies.
First, the attacker may continue sending corrupted  inputs to the compromised
compartment after recovery is triggered.
As discussed earlier, the surrogate is unlikely to have
the same memory corruption vulnerabilities as the original compartment,
and it is unlikely to be exploitable in the same manner.
Although corrupt inputs may cause the surrogate to produce degraded outputs,
the attacker can no longer hijack the control flow of the system,
reducing the attack impact from a full system compromise and crash to slight mission degradation
(i.e., landing with an acceptable deviation to destination points as shown in \S\ref{subsec:evaluation-effectiveness}).
Second, the attacker may attempt to evade CFI/DFI detection entirely so that
recovery is never triggered.
As discussed in the \textit{Detection Techniques},
this is a limitation of the underlying detection mechanism. 
Third, the attacker may attempt to tamper with the trained surrogate model.
Since surrogates are trained offline prior to deployment, 
this would require the attacker modifying the training environment, which is outside our threat model. 
Further, we assume that the surrogates are trustworthy and that modifying them would require root privileges or physical access, both of which fall outside our threat model (\S\ref{subsec-assumption}).
Finally, ML attacks, such as adversarial attacks, and their corresponding defense techniques constitute a separate line of research and are orthogonal to this work.


\label{discussionsec}

\section{Related work}

\noindent\textbf{Attack Detection.}
Various techniques have been proposed to detect memory corruption attacks, 
including stack protection~\cite{cowan1998stackguard}, 
control flow integrity (CFI)~\cite{abadi2009control,davi2015hafix}, 
data flow integrity (DFI)~\cite{castro2006securing}, 
memory protection~\cite{song2016enforcing,salehi2026rvdebloater,kim2018securing,abubakar2021shard}, 
hardware-based methods~\cite{koeberl2014trustlite,ziad2021no,nyman2017cfi,zhou2014armlock},   
and compartmentalization-based approaches~\cite{lefeuvre2025sok,clements2018aces} that apply the least-privilege principle. 

Mini-Me~\cite{ding2021mini} employs an LSTM-based model to predict 
the RV controller's expected state outputs, 
and triggers a fail-safe mechanism when the difference between the prediction and actual measurements exceeds a threshold, 
thereby detecting malicious physical control-state modifications caused by external sensor spoofing and internal control-parameter tampering.
However, it focuses solely on detection and, upon detecting an attack, terminates the process and triggers the fail-safe mechanism without providing recovery capability.
Further, unlike our work, Mini-Me focuses on the controller function and cannot be directly applied to other parts of the firmware, 
which must also be considered for effective recovery. Extending this approach to train a separate ML model for every firmware function 
would also be impractical for CPSs.


\noindent\textbf{Attack Recovery.}
Recovery techniques for CPSs can be categorized into three types: redundancy-based techniques, physical attack recovery, and memory corruption attack recovery.

(1) \textbf{\textit{Redundancy.}}
Redundancy-based techniques are primarily designed for fault tolerance, enabling recovery by replacing faulty components with identical backup instances that execute concurrently, often involving system reinitialization. 
These techniques can be broadly categorized into hardware-based~\cite{fei2018cross,siewiorek2005fault,aggarwal2007configurable,qin2025mvtee} 
and software-based approaches~\cite{ron2025galapagos,rosti2024ll,forrest1997building}.
Both types of techniques require either replicating physical components (e.g., processors or sensors) or maintaining multiple instances of an entire program, such as N-version programming~\cite{avizienis1985n}, switching to a backup upon failure.
However, both incur significant overhead, either in hardware cost and design complexity or in concurrent execution, making them unsuitable for resource-constrained CPSs, including RVs.
Furthermore, since all instances are typically implemented using similar representations, they likely have the same memory corruption vulnerabilities
and remain exploitable by an attacker, making these techniques ineffective for attack recovery. 

Recovery Blocks~\cite{randell1975} (RBs) is a classic technique for fault-tolerance where an alternate implementation is invoked when an acceptance test on the primary's output fails. While conceptually similar, \technique{} differs from RBs in two ways. First, alternates have to be manually implemented in RBs, whereas we learn surrogates automatically using ML. Second, RBs do not target security, and hence the alternates, being conventional implementations, remain susceptible to the same memory corruption vulnerabilities as the primary. \technique{}, in contrast, learns ML based surrogates that have a different representation from the original, and hence are unlikely to contain the same memory corruption vulnerabilities.

(2) \textbf{\textit{Physical Recovery.}}  
Many recovery techniques for physical attacks in RVs have been proposed~\cite{jeong2023rocking,fei2020learn,choi2020software,dash2021pid,dash2024specguard,park2023scvmon}.
These techniques leverage two different approaches to enable recovery.
First, some techniques~\cite{jeong2023rocking,choi2020software} use software-based pre-trained sensor models to replace
the hardware sensors under attack and correct their measurements.
Second, since relying solely on sensor measurements is not sufficient,
other techniques~\cite{dash2021pid,dash2024specguard,fei2020learn} train the RV's physical controller (i.e., PID) as a recovery controller.
However, as mentioned in \S\ref{sub-sec:current-limitations}, they focus on physical attacks in RVs 
and replace only the physical sensors
or the PID controller upon attack detection, which is not effective for memory corruption attacks.

(3) \textbf{\textit{Memory Corruption Recovery.}}
Different recovery techniques have been proposed 
 including software rejuvenation~\cite{huang1995software}, failure-oblivious computing~\cite{rinard2004enhancing}, and micro-rebooting~\cite{candea2004microreboot}. 
However, these techniques require rebooting the entire system or program each time an attack occurs, 
leading to prolonged unavailability~\cite{qin2005rx}. 
Micro-rebooting techniques attempt to address this issue by rebooting only the affected part of the program or system, 
but it still incurs significant overhead and, once rebooting is complete, that part remains susceptible to exploitation by attackers~\cite{li2025software}.
Failure-oblivious computing techniques attempt to prevent the program from crashing upon buffer overflow (e.g., out-of-bound reads) 
by returning artificial random values; however, they incorrectly alter the system's behavior and introduce silent errors that may not manifest immediately, 
potentially leading to incorrect system operation~\cite{qin2005rx}.

The Simplex approach~\cite{bak2009system} outlines a system that switches to a simpler instance manually written by the developers upon detecting a fault. Inspired by this approach,  
Li et al.~\cite{li2025software} proposed Gecko, 
which leverages software compartmentalization introduced in ACES~\cite{clements2018aces}
to partition firmware at the task level into individual compartments. 
Upon attack detection, it replaces the compromised compartment with either an automated version that returns pre-recorded default values captured 
during the device initialization phase, or a manually written custom version. 
In our evaluation, \technique{} successfuly recovers all five missions under attacks, whereas Gecko recovered none of them (\S\ref{subsec:evaluation-effectiveness}), demonstrating that the default value version is insufficient for recovery.
On the other hand, writing a custom version of the compartment for recovery demands
substantial manual effort and 
 system knowledge.  


\noindent\textbf{Accelerated Substitution.}
Techniques have been proposed to improve system performance by replacing parts of an application 
with surrogate ML  models~\cite{schutt2017schnet,sadowski2016synergies,dong2023auto,esmaeilzadeh2012neural,moreau2015snnap}.
For example, ML surrogate models can be optimized through design choices 
such as the number of neural network layers, to trade off prediction accuracy and computational cost.
These techniques inspired our design of \technique{}. 
However, they are designed for performance optimization 
rather than security or attack recovery. 
Further, since their goal is  performance acceleration, 
they only choose the most time-consuming code regions, 
train a surrogate to approximate the output, and leave the other parts of the application unchanged. 
\label{releatedsec}

\section{Conclusion}
We propose \technique{}, a framework for automatically 
recovering CPSs from memory corruption attacks using ML-based 
surrogates. 
\technique{} works at the compartment granularity by \textit{automatically} generating surrogates using ML that closely 
approximate their original compartments' behavior without being susceptible to the same memory corruption attack. 
At runtime, upon attack detection, \technique{} 
replaces the compromised compartment with its trained surrogate.  
\technique{} is implemented using the LLVM compiler, making it portable.

We evaluated \technique{} on seven different RVs, 
including four simulated and three real ones, 
across diverse scenarios and real-world attacks.
Our results show that \technique{} generates surrogates 
with an average R$^2$ of 96\% 
relative to the original compartments, 
successfully recovers the system and completes the tasks 
despite attacks, 
and incurs an average CPU overhead of 8.5\% and memory overhead of 9\%.
\label{conclusionsec}

\cleardoublepage


\bibliographystyle{plainurl}
\bibliography{main}

\appendix
\section{Memory Corruptions}
\label{Sec:Back-memory}


Embedded systems software, including RV firmware, is typically written in C and C++ 
because they enable low-level hardware control and efficient execution.
Unfortunately, these languages provide no built-in memory safety,  
resulting in memory corruption vulnerabilities such as buffer overflows.  
Attackers can exploit such vulnerabilities to manipulate device behavior or take full control of the system's execution flow~\cite{szekeres2013sok}, e.g., 
control-flow hijacking attacks.


Control flow hijacking attacks occur when a code pointer, 
such as a function pointer (forward edge) or a return address (backward edge),
becomes corrupted. 
By hijacking the control flow, attackers can either execute malicious payloads injected
into the application stack~\cite{szekeres2013sok} (\textit{code injection attacks}) or reuse existing code sequences (gadgets) 
from the victim application~\cite{shacham2007geometry} (\textit{code reuse attacks}). 
Security mechanisms like Data Execution Prevention (DEP), 
which marks memory regions as either writable or executable (i.e., $W \oplus X$), mitigate code injection attacks. 
However, code reuse attacks bypass these defenses by injecting addresses of existing instructions 
into corrupted code pointers, altering the original control flow in an arbitrarily expressive way. 
\textit{Code reuse attacks} are categorized into \textit{Return-Oriented Programming} (ROP)~\cite{shacham2007geometry}
and \textit{Jump-Oriented Programming} (JOP)~\cite{bletsch2011jump}, based on whether the corrupted pointers target return instructions 
or function pointers, respectively. 

\section{Ablation Study}
\label{appendix_sec:ablation}

\begin{figure}[h]
    \centering
    \includegraphics[width=0.72\linewidth]{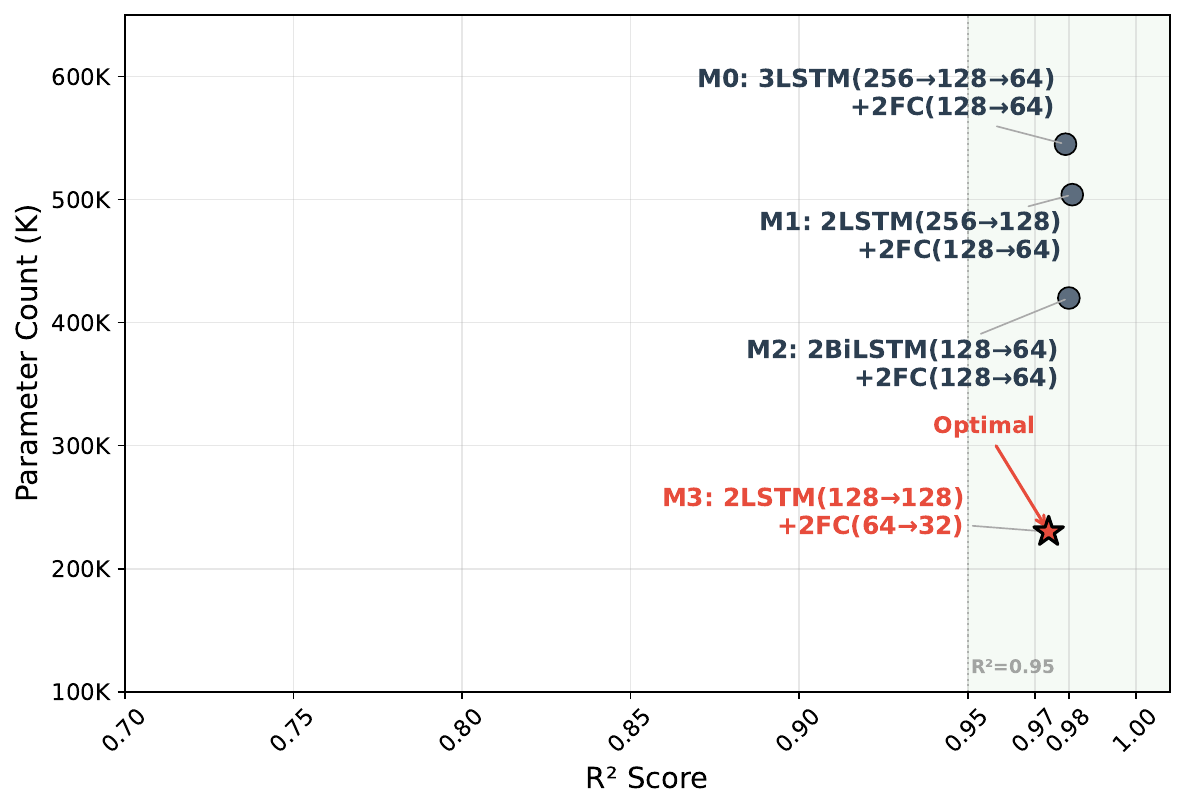}
    \caption{Predictive performance-overhead trade-off across surrogate model architectures.
    Each point represents a candidate architecture evaluated on R$^2$
    (higher is better) and parameter count (lower is better). M3
    achieves competitive R$^2$=0.974 while requiring more than
    $1.8\times$ fewer parameters than all other top-performing architectures.}
    \label{fig:ablation}
\end{figure}

Figure~\ref{fig:ablation} shows the results for the four most
representative candidate architectures. 
M0 uses three stacked
LSTM layers with widths 256, 128, and 64 followed by two
fully-connected layers of 128 and 64 units, totaling 545K
parameters. M1 reduces depth to two LSTM layers of widths 256 and
128 with two fully-connected layers of 128 and 64 units (504K
parameters). M2 replaces standard LSTM layers with bidirectional
LSTM layers of widths 128 and 64 followed by two fully-connected
layers (420K parameters), while M3 (our selected design) uses two
LSTM layers both of width 128 followed by two fully-connected
layers of 64 and 32 units (230K parameters). 

M3 achieves R$^2$=0.974, representing the best predictive performance-overhead trade-off, as M0, M1, and M2, despite marginally higher R$^2$ scores of 0.979, 0.981, and 0.980, respectively, each require more than $1.8\times$ more parameters. Furthermore,
M2's bidirectional design incurs higher memory overhead since
bidirectional LSTMs maintain hidden states for both forward and
backward passes, doubling the memory overhead at runtime.
Therefore, we employ M3 as the surrogate architecture for all
compartments across all subject RVs in our experiments.

\end{document}